\documentclass[3p,times,twocolumn]{elsarticle}

\usepackage{ecrc}
\usepackage{amsmath}

\volume{00}

\firstpage{1}

\journalname{Nuclear Physics B Proceedings Supplement}

\runauth{N. Carrasco}

\jid{nuphbp}

\jnltitlelogo{Nuclear Physics B Proceedings Supplement}

\usepackage{amssymb}

\usepackage[figuresright]{rotating}

\begin{document}

\begin{frontmatter}



\dochead{}

\title{Neutral meson oscillations on the lattice}


\author{N. Carrasco}

\address{INFN - Sezione di Roma Tre}

\begin{abstract}
Accurate measurements of K, D and B meson mixing amplitudes provide stringent
constraints in the Unitary Triangle analysis, as well as useful bounds on New Physics
scales. 
Lattice QCD provides a non perturbative tool to compute the hadronic matrix
elements entering in the effective weak Hamiltonian, with errors at a few percent
level and systematic uncertainties under control. 
I  review  recent lattice results for these hadronic matrix element performed with $N_f=2$, $N_f=2+1$ and $N_f=2+1+1$ dynamical sea quarks.
\end{abstract}


\end{frontmatter}


\section{Introduction}
\label{}
The mixing of neutral pseudoscalar mesons plays an important role in the understanding of the physics of CP-violation in the Standard Model (SM) and in providing important constraints on theories Beyond Standard Model (BSM). In the SM, the effective weak Hamiltonian of neutral meson oscillations receives contribution only from the vector-axial local operator 
\begin{equation}
Q_{1}=\left[\bar{h}^{a}\gamma_{\mu}(1-\gamma_{5})q^{a}\right]\left[\bar{h}^{b}\gamma_{\mu}(1-\gamma_{5})q^{b}\right]
\end{equation}
where $h=b,c,s$ denotes a bottom, charm or strange quark and $q=d,s,u$ denotes a light quark (down, strange or up).
For neutral $K$ and $B_{(s)}$ mesons the hadronic matrix element of $Q_1$ is the dominant contribution for SM predictions of the corresponding mixing observables, $\epsilon_K$ and $\Delta M_{d(s)}$ respectively.

The experimental observable quantity $\epsilon_K$  is related to the matrix element of $Q_1$ between a kaon and an anti-kaon state through 
\begin{equation}
\begin{array}{l}
\label{eq:epsilon_K}
\epsilon_K=  \kappa_{\epsilon}  \dfrac{G_F^2 f_K^2 M_K M_W^2}{6 \sqrt{2} \pi^2 \Delta M_K } \hat{B}_K |V_{cb}|^2 |V_{us}|^2 \bar{\eta}\\
  \!\times \!\left[-\eta_1 S_{cc}\! \left( 1-\lambda^2/2\right )\! +\! \eta_2 S_{tt}|V_{cb}|^2\lambda^2 \!(1-\bar{\rho})\! +\! \eta_3 S_{ct} \right]\!

\end{array}
\end{equation}
where $S_{cc}$, $S_{tt}$ and $S_{ct}$ are the Inami-Lim functions giving the charm, top and charm-top contributions to the box diagram and $\eta_i$ contain the short distance QCD contributions. Explicit expressions for $\eta_i$ at leading order can be found in \cite{Buras:1990fn,Herrlich:1993yv,Herrlich:1996vf}. Recently, $\eta_1$ and $\eta_3$ have been calculated at NLL \cite{Brod:2010mj,Brod:2011ty}.  The Renormalization Group Invariant (RGI) $\hat{B}_K$ parameter in Eq.(\ref{eq:epsilon_K}) is defined at NLL by
\begin{equation}
\hat{B}_K=\left(\dfrac{\alpha_s(\mu)}{4\pi}\right)^{-\gamma_0/{2\beta_0}} \left[1+\dfrac{\alpha_s(\mu)}{4\pi} J \right] B_K(\mu)
\end{equation}
where $\gamma_0$ and $J$ are the anomalous dimension at one and two loops \cite{Ciuchini:1997bw}, respectively. The $B_K(\mu)$ parameter parametrizes the deviation of the hadronic  matrix element from the Vaccum Insertion Approximation (VIA) estimate
\begin{equation}
\langle \overline{K}^0 | Q_1 (\mu) | K^0 \rangle \!\! \equiv\!\! \langle \overline{K}^{0}|Q_{1}(\mu)|K^{0}\rangle_{\textrm{VIA}} B_K (\mu)\!\! =\!\! \dfrac{8}{3} f_K^2M_K^2 B_K(\mu)
\end{equation}

Eq.(\ref{eq:epsilon_K}) defines an hyperbola in the $(\bar{\rho},\bar{\eta})$ plane of the Unitary Triangle which can be used to (over)-constrain the upper vertex. 

In the case $B$-mesons, the SM prediction for the mass difference of neutral $B_q$-meson, $q=s,d$, is given by
\begin{equation}
\Delta M_{Bq} = \dfrac {G_F^2 M_W^2}{6\pi^2} \eta_B S_t M_{Bq} |V_{tb}V_{tq}^*|^2 f_{Bq}^2 \hat{B}_{Bq}
\end{equation}
It is  interesting to consider the SU(3)-breaking ratio $\xi$ defined by
\begin{equation}
\xi=\dfrac{f_{Bs}\sqrt{B_{Bs}}}{f_{Bd}\sqrt{B_{Bd}}}
\end{equation}
Its value combined with the experimental measurement of the mass differences for the $B_{Bs}$ and $B_{Bd}$ mesons determine the ratio $|V_{ts}/V_{td}|$, which provides a constraint in the UT
\begin{equation}
\dfrac{\Delta M_{Bs}}{\Delta M_{Bd}} = \left| \dfrac{V_{ts}}{V_{td}}\right|^2 \dfrac{M_{Bs}}{M_{Bd}} \xi^2
\end{equation}

For the D-meson mixing long-distance effects coming from the insertion of two effective $\Delta C=1$ operators dominate over the SM short distance contributions. However, it is still possible to put significant constraints on the New Physics (NP) parameter space and discriminate between BSM theories by considering the complete basis of local $\Delta C=2$ operators.

In BSM theories, addition local four-fermion operators can contribute. The most general $\Delta F=2$ effective hamiltonian, with $F=S,C,B$, can be written in terms of five operators
\begin{equation}
H_{\textrm{eff}}^{\Delta F=2}={\displaystyle \sum_{i=1}^{5}C_{i}(\mu)Q_{i}(\mu)},
\label{eq:Heff}
\end{equation}
where $C_i$ are the Wilson coefficients which encode the short distance contributions and $\mu$ is the renormalization scale. The operators $Q_i$, in the so-called SUSY basis, are\footnote{The complete basis contains three additional operators $\tilde{Q}_{1-3}$ obtained from $Q_{1-3}$ with the exchange of $(1-\gamma_5)\leftrightarrow (1+\gamma_5)$.  The parity-even parts of the operators  $\tilde{Q}_i$, which is the only one contributing in strong interactions, coincide with those of the operators $Q_i$. Therefore, only the operators $Q_i$ need to be considered.}
\begin{equation}
\begin{array}{l}
Q_{1}=\left[\bar{h}^{a}\gamma_{\mu}(1-\gamma_{5})q^{a}\right]\left[\bar{h}^{b}\gamma_{\mu}(1-\gamma_{5})q^{b}\right]\\
Q_{2}=\left[\bar{h}^{a}(1-\gamma_{5})q^{a}\right]\left[\bar{h}^{b}(1-\gamma_{5})q^{b}\right] \\
Q_{3}=\left[\bar{h}^{a}(1-\gamma_{5})q^{b}\right]\left[\bar{h}^{b}(1-\gamma_{5})q^{a}\right]\\
Q_{4}=\left[\bar{h}^{a}(1-\gamma_{5})q^{a}\right]\left[\bar{h}^{b}(1+\gamma_{5})q^{b}\right] \\
Q_{5}=\left[\bar{h}^{a}(1-\gamma_{5})q^{b}\right]\left[\bar{h}^{b}(1+\gamma_{5})q^{a}\right].
\label{eq:Qi}
\end{array} 
\end{equation}
The long-distance contributions are described by the matrix elements of the renormalized four-fermion operators. The renormalized bag parameters, $B_i$ $(i=1,...,5)$, provide the value of four-fermion matrix elements in units of the deviation from their vacuum insertion approximation. They are defined as
\begin{equation}
\begin{array}{l}
\langle\overline{P}^{0}|Q_{1}(\mu)|P^{0}\rangle=C_{1} B_{1} (\mu)m_{P}^{2}f_{P}^{2} ,\\
\langle\overline{P}^{0}|Q_{i}(\mu)|P^{0}\rangle=C_{i} B_{i}(\mu) m_{P}^{2}f_{P}^{2} \dfrac{m_{P}^{2}}{\left(m_{h}(\mu)+m_{q}(\mu)\right)^{2}} ,
\label{eq:Bi-def}
\end{array}
\end{equation}
where $C_i={8/3,-5/3,1/3,2,2/3}$, $i=1,..,5$. $|P^0\rangle$ is the pseudoscalar, $K$, $D$ or $B$ state, $m_{P}$ and $f_{P}$ are the pseudoscalar mass and decay constant and $m_{h}$ and $m_{q}$ are the renormalized quark masses. 

In NP analysis, the computation of the relevant matrix elements combined with the experimental observables offers the chance to constraint the parameters appearing in NP models. These constraints can be obtained in a model-independent way as presented in \cite{Bertone:2012cu,Carrasco:2014uya,Carrasco:2013zta}. 

\subsection{$B_i$-parameter calculation on the lattice}
$B_i$-parameters on the lattice are evaluated from a three-point correlation function with the insertion of two pseudoscalar meson fields at two time slices, $t_1$ and $t_2$, and inserting the $\Delta F=2$ four-fermion operator at any time slice $t_0$ with $t_0 \in [t_1,t_2]$.
The estimate of $B_i$ is obtained for large time separation, i.e $t_1 \ll t_0 \ll t_2$
\begin{equation}
\begin{array}{lll}
\!\!\dfrac{\langle\mathcal{P}(t_{1})Q_{1}(t_0)\mathcal{P}^{\dagger}(t_{2})\rangle}{C_1\langle\mathcal{P}(t_{1})A(t_0)\rangle\langle A^{\dagger}(t_0)\mathcal{P}(t_{2})\rangle} & \!\!\!\!\!\!\xrightarrow[t_{1}\ll t_0\ll t_{2}]{}\!\!\!\!\!&\dfrac{\langle\overline{P}^{0}|Q_{1}(\mu)|P^{0}\rangle}{C_{1} \langle\overline{P}^{0}|A_0|0\rangle \langle 0 | A^{\dagger}_0|P^0\rangle} \\
\!\!& &= B_{1} \\
\!\!\dfrac{\langle\mathcal{P}(t_{1})Q_{i}(t_0)\mathcal{P}^{\dagger}(t_{2})\rangle}{C_i\langle\mathcal{P}(t_{1})P(t_0)\rangle\langle P(t_0)\mathcal{P}(t_{2})\rangle} &  \!\!\!\!\!\!\xrightarrow[t_{1}\ll t_0\ll t_{2}]{}\!\!\!\!\!& \dfrac{\langle\overline{P}^{0}|Q_{1}(\mu)|P^{0}\rangle}{C_{i} \langle\overline{P}^{0}|P_0|0\rangle \langle 0 | P^{\dagger}_0|P^0\rangle} \\
\!\!& & = B_{i}\,\,\, i > 2\\
\end{array}
\label{Bi-latt}
\end{equation} 
where $\mathcal P$ creates the pseudoscalar meson and $P$ ($A$) is the pseudoscalar (axial) current which satisfy  $ \langle 0 | P^{\dagger}_0|P^0\rangle=f_P M_P / (m_h+m_q)$ and $ \langle 0 | A^{\dagger}_0|P^0\rangle=f_P M_P$, respectively.

A renormalization step is necessary to get the results in the continuum limit. The four-fermion renormalization constants for the operators $Q_i$ are the links between the matrix elements computed on the lattice and the ones computed on the continuum.

On the lattice, with Wilson fermions, due to the  chiral symmetry breaking,  four-fermion opearators mix with additional dimension-six four-fermion operators which belong to a different representation of the chiral group. 

The European Twisted Mass Collaboration (ETMC) bypass this complication with the use of a mixed action setup adopting different regularizations for the sea  and the valence quarks. In particular, as it was proposed in \cite{Frezzotti:2004wz}, they introduce the Twisted Mass action for the sea quarks while on the valence the Osterwalder-Seiler action, a variant of the Twisted Mass action, is implemented. This strategy provides a computation  framework without wrong chirality mixing effects and free of $\mathcal{O}(a)$ discretization errors.

Staggered fermions retain a remnant chiral symmetry but at non zero lattice spacing the taste symmetry is broken. The mixing with other dimension-six four-fermion operators due to the broken taste symmetry is usually treated via Staggered Chiral Perturbation Theory ($S{\chi}PT$).  

Domain wall fermions also preserves chiral symmetry. However, residucal chiral symmetry breaking due to the finite extent of the 5th dimension induces mixing with dimension six operators, which in practice  turns out to be negligibly small \cite{Aoki:2007xm}.

\vspace*{-0.5cm}
\section{Recent lattice $B_K$ computations}

A great deal of effort has been devoted to compute $B_K$ by different lattice collaborations using different actions with well controlled and small $\mathcal{O}(a^2)$ discretization errors and with light quark masses in the chiral regime.
Recently, five collaboration have computed the continuum limit value of $B_K$ performing unquenched simulations. 

\vspace*{-0.4cm}
\begin{figure}[h!]
 \centering
 \vspace*{-3.0cm}
 \hspace*{-3.6cm}\includegraphics[scale=0.65]{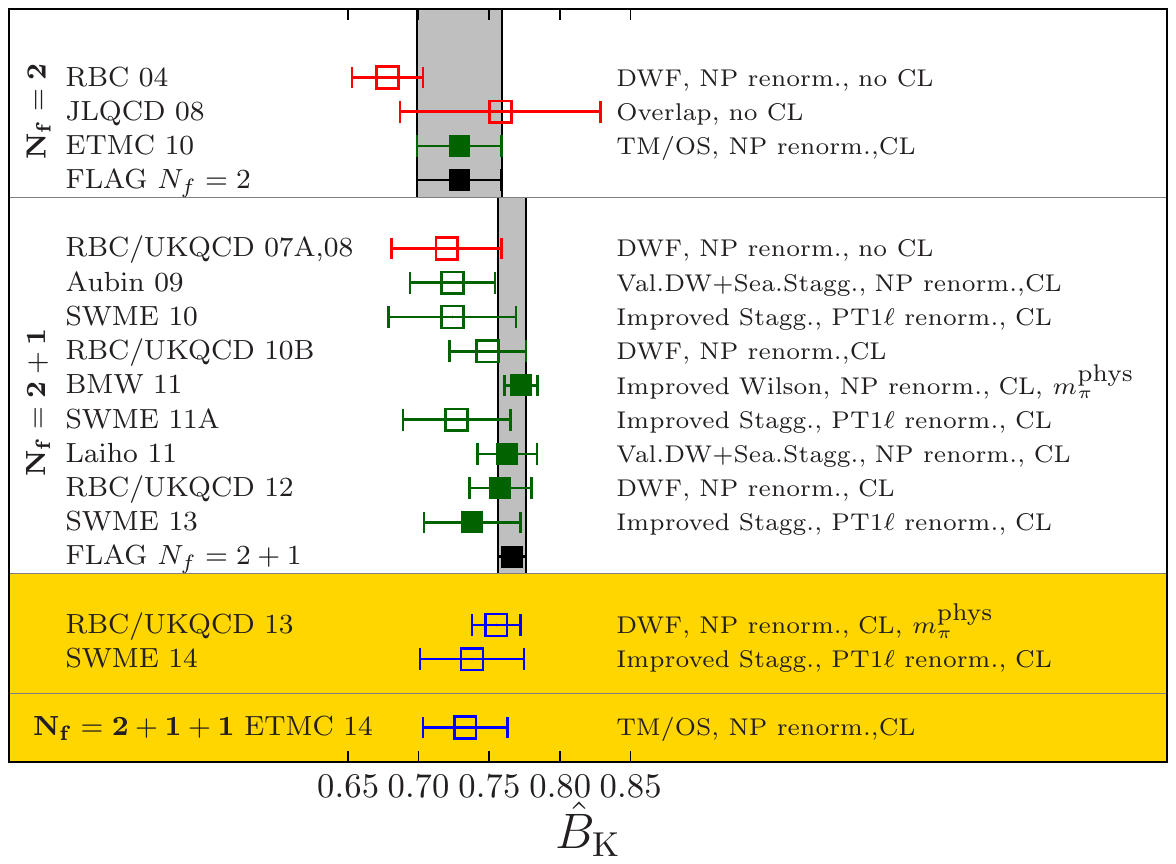}
 \vspace*{-11.2cm}

 \caption{ \label{fig:BK}Overview of lattice results for the $B_K$ parameter. Points with white background are those included in the FLAG report while those with yellow background are new. The black squares and grey bands indicate the FLAG global averages while the significance of the point colours with white background follows the FLAG colour code. }
\end{figure}

\vspace*{-0.6cm}
\begin{table}[h!]
\center
\begin{footnotesize}
\begin{tabular}{lll}
\hline 
Collaboration  \hspace*{-0.35cm} &   $N_{f}$ &  $\hat{B}_{K}$\tabularnewline
\hline 
ETMC \cite{Bertone:2012cu} \hspace*{-.4cm}  & 2 & \hspace*{-0.3cm}  $0.729(25)(17)$\tabularnewline
BMW \cite{Durr:2011ap} \hspace*{-.4cm}  & 2+1 & \hspace*{-0.35cm}  $0.7727(81)(84)$\tabularnewline
SWME \cite{Bae:2014sja} \hspace*{-.4cm}  & 2+1 & \hspace*{-0.3cm}  $0.7379(47)(365)$\tabularnewline
Laiho \cite{Laiho:2011np}  \hspace*{-.4cm}  & 2+1 & \hspace*{-0.3cm}  $0.7628(28)(205)$\tabularnewline
RBC \cite{Arthur:2012opa}  \hspace*{-.4cm}  & 2+1 & \hspace*{-0.3cm}  $0.755(4)(15)$\tabularnewline
ETMC \cite{tobepublished} \hspace*{-.4cm}  & 2+1+1 & \hspace*{-0.3cm}  $0.730(17)(15)$\tabularnewline
FLAG \cite{Aoki:2013ldr}\hspace*{-.4cm}  & 2 & \hspace*{-0.3cm}  $0.729(25)(17)$ \tabularnewline
FLAG  \cite{Aoki:2013ldr}\hspace*{-.4cm}  & 2+1 & \hspace*{-0.3cm} $0.7661(99)$  \tabularnewline
\hline
\end{tabular}
\end{footnotesize}
\caption{\label{tab:BK}Recent lattice results for $\hat{B}_K$ by the various collaborations. First error is statistical while the second one is systematic.}
\end{table}

Fig.\ref{fig:BK} collects the  $\hat{B}_K$ results published in the last five years by the various collaborations together with the FLAG averages \cite{Aoki:2013ldr}. Table \ref{tab:BK} reports the most recent numbers published by each collaboration and the FLAG averages. 


The BMW collaboration presented the  first $\hat{B}_K$ result at the physical point  \cite{Durr:2011ap}. Simulations are performed using the HEX-smeared clover-improved Wilson  action with $N_f=2+1$ dynamical sea quarks and four lattice spacing in the range $0.054\, \mbox{fm} \leq a \leq 0.093\, \mbox{fm}$. Non perturbative renormalization is performed in the RI-MOM scheme and they find that the smearing of the link variables reduces the operator mixing induced by chiral symmetry breaking. The BMW $B_K$ result is the most precise result to date with a total uncertainty of 1.5\% dominated by the truncation of the perturbative NLL matching from RI-MOM to RGI or $\overline{MS}$. 

The SWME  \citep{Bae:2014sja} uses improved staggered valence quarks on the staggered  $N_f=2+1$ AsqTad MILC ensembles. They use four lattice spacings from 0.12 fm to 0.045 fm and  minimum pion mass of 174 MeV. 
The main improvement in \cite{Bae:2014sja} with respect \cite{Bae:2011ff} is the inclusion of  additional ensembles which allow a controlled extrapolation in the quark masses.
Taste breaking, which is reduced thanks to the smearing techniques, can be incorporated in the the chiral and continuum extrapolations based on $S{\chi}PT$ ansatze. The dominant error comes from the use of one-loop perturbative renormalization ($\sim 4\%$).  

The Laiho and Van der Water result \cite{Laiho:2011np}  uses a mixed action: valence domain wall fermions over the staggered $N_f=2+1$ AsqTad MILC ensembles.Three lattice spacings ranging from 0.12 fm to 0.06fm and pion masses as low as 210 MeV are used in the simulation. Since domain wall valence quarks are used, mixing with wrong chirality operators only occurs due to small residual chiral symmetry breaking on the lattice.  The main improvement with respect to its previous determination \cite{Aubin:2009jh} is the implementation of the RI-SMOM non perturbative renormalization scheme which suppress chiral symmetry breaking and other infrared effects and reduces the size of the one-loop perturbative coefficient responsible for the matching to the continuum $\overline{MS}$ scheme.  The largest error they report is due to the perturbative matching between the lattice and the continuum.

The RBC/UKQCD result in \cite{Arthur:2012opa} is obtained with $N_f=2+1$  domain wall fermions both in the valence and in the sea at three lattice spacings in the range $[0.09\!\!:\!\!0.14]$ fm. The renormalization of the four-fermion operator is performed with two RI-SMOM variants where non-exceptional momentum renormalization conditions and twisted boundary conditions are applied. They report as dominant source of error the perturbative truncation in the matching to the continuum scheme. Preliminary results with physical mass ensembles are reported in the Lattice 2013 \cite{Frison:2013fga} and Lattice 2014 conferences. 

ETMC adopts a mixed action setup with maximally twisted sea quarks and Osterwalder-Seiler valence quarks. ETMC has carried out the computation working with $N_f=2$ dynamical sea quarks, four lattice spacing in the range $[0.05\!:\!0.1]$ fm and pion masses as low as 270 MeV \cite{Constantinou:2010qv,Bertone:2012cu}. Preliminary results with $N_f=2+1+1$ dynamical sea quarks have been presented in the Lattice 2013 conference \cite{Carrasco:2013jaa}.

\vspace*{-0.5cm}
\section{$K^0-\overline{K}^0$  mixing beyond the Standard Model}

There is a considerable recent activity concerning the matrix elements of the complete basis of four-fermion operators controlling the $K^0-\overline{K}^0$ mixing. Three results have been presented recently, two of them with $N_f=2+1$. 

The SWME collaboration has published results using improved staggered fermions  \cite{Bae:2013tca}. $B_i$ estimators are extrapolated to the continuum from three lattice spacing ranging down to $a\sim0.045$ fm. As in the $B_K$ publication \cite{Bae:2014sja}, renormalization is performed perturbatively at one loop.  Compared to other lattice computations, one particularity is the construction of "golden combinations" for the chiral extrapolations which cancel the chiral logarithms at NLL. 
In  \cite{Boyle:2012qb} the RBC/UKQCD computation using Domain Wall fermions on a single lattice spacing is reported. An update with two lattice spacings is presented in \cite{Lytle:2013oqa} and in the Lattice 2014 conference.
Although the renormalization is performed in several intermediate schemes, including non-exceptional schemes that avoid unwanted infrared effects, only final results obtained from the RI-MOM scheme are quoted since only in this case the conversion factors to $\overline{MS}$ are available for the complete set of operators. 

Finally, the ETMC has published final continuum-limit results for the complete basis of $Q_i$ operators using  $N_f=2$ ensembles \cite{Bertone:2012cu} as well as preliminary results with  $N_f=2+1+1$ ensembles \cite{Carrasco:2013jaa}. The computational setup is  the same as in the $B_K$ computation and the four-fermion renormalization constants are computed non perturbatively in the RI-MOM scheme and converted perturbatively to $\overline{MS}$. 
\begin{table}[h!]
\center
\begin{footnotesize}
 \begin{tabular}{llcccc}
\hline 
 Collaboration \hspace*{-.4cm} & $N_f$ \hspace*{-.3cm} & $B_{2}$\hspace*{-.25cm} & $B_{3}$ \hspace*{-.25cm}& $B_{4}$\hspace*{-.25cm} & $B_{5}$\hspace*{-.25cm}\tabularnewline
 \hline 
ETMC \cite{Bertone:2012cu}\hspace*{-.4cm} & 2 \hspace*{-.3cm} & 0.47(2) \hspace*{-.25cm}& 0.78(4)\hspace*{-.25cm} & 0.76(3) \hspace*{-.25cm}& 0.58(3)\hspace*{-.25cm}\tabularnewline
RBC \cite{Boyle:2012qb} \hspace*{-.4cm} & 2+1\hspace*{-.3cm}  & 0.43(5)\hspace*{-.25cm} & 0.75(9)\hspace*{-.25cm} & 0.69(7) \hspace*{-.25cm}& 0.47(6)\hspace*{-.25cm}\tabularnewline 
 SWME  \cite{Bae:2013tca} \hspace*{-.4cm} & 2+1 \hspace*{-.3cm} & 0.55(3) \hspace*{-.25cm}& 0.79(3)\hspace*{-.25cm} & 1.03(5) \hspace*{-.25cm}& 0.86(4)\hspace*{-.25cm}\tabularnewline 
 ETMC \cite{tobepublished} \hspace*{-.4cm} & 2+1+1 \hspace*{-.3cm} & 0.46(2)\hspace*{-.25cm} & 0.79(4)\hspace*{-.25cm} & 0.77(4)\hspace*{-.25cm} & 0.48(4)\hspace*{-.25cm}\tabularnewline
 \hline 
\end{tabular}
\end{footnotesize}
\caption{\label{tab:K-K}$B_{2-5}$ for K-mixing renormalized  in $\overline{MS}$ scheme of \cite{Buras:2000if} at 3 GeV obtained by ETMC, RBC/UKQCD and SWME.}
\end{table}
 
\begin{figure}[h!]
\vspace*{-1.7cm}
\begin{tabular}{cc}
\hspace*{-2.3cm}\includegraphics[scale=0.34]{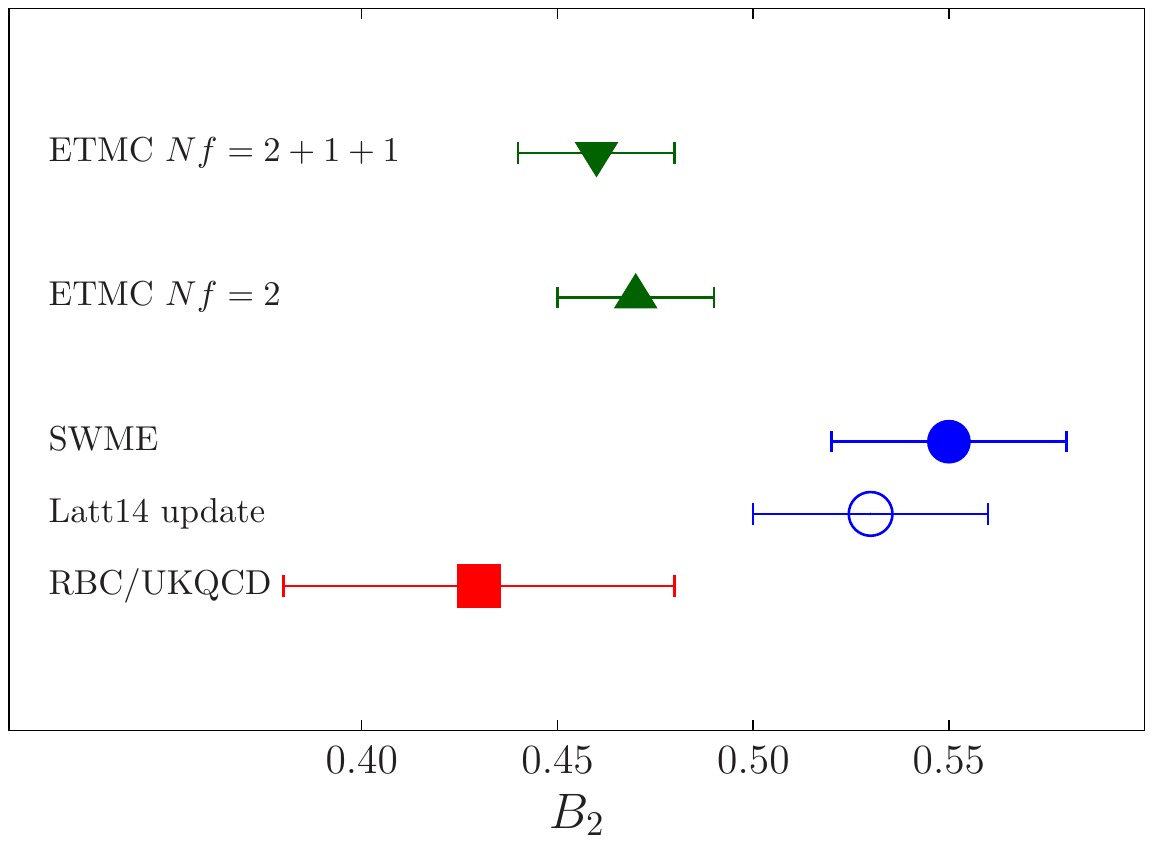}\vspace*{-3.8cm} & \hspace*{-3.5cm}\includegraphics[scale
=0.34]{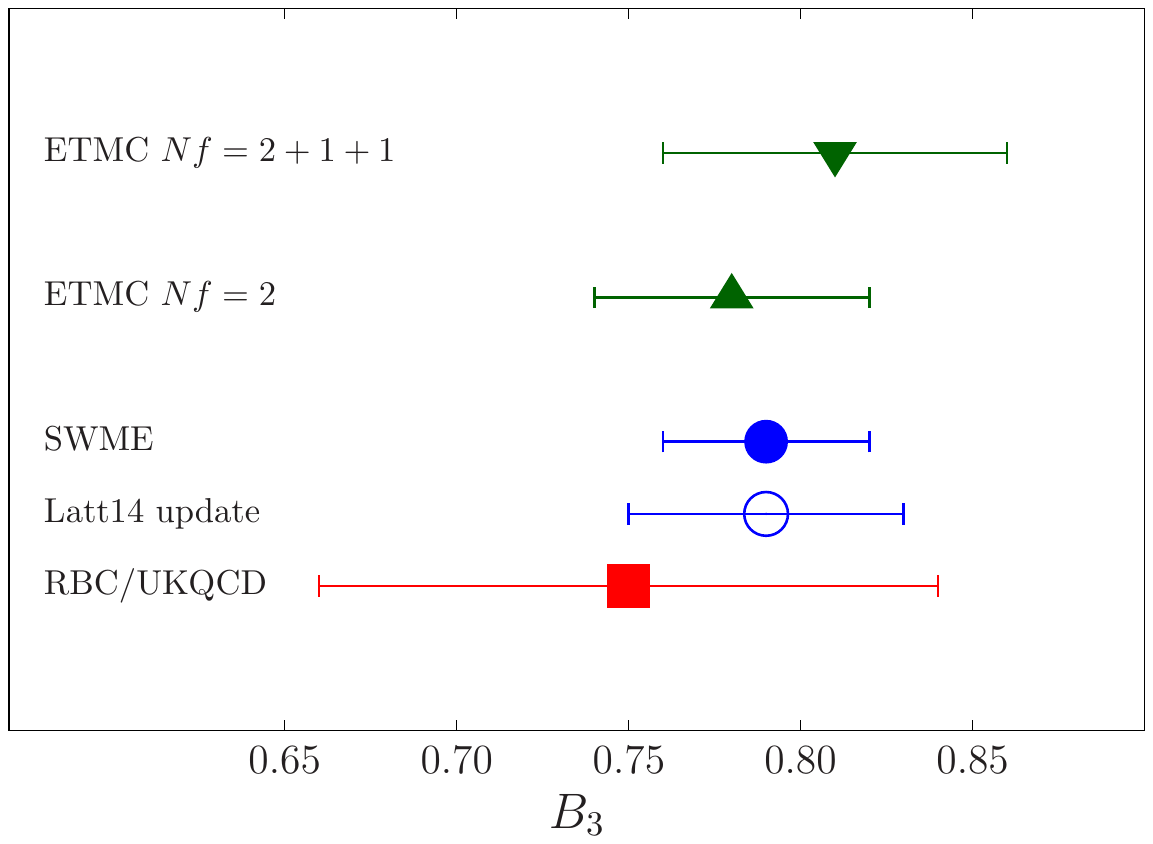}\vspace*{-3.2cm}\tabularnewline
\hspace*{-2.3cm}\includegraphics[scale=0.34]{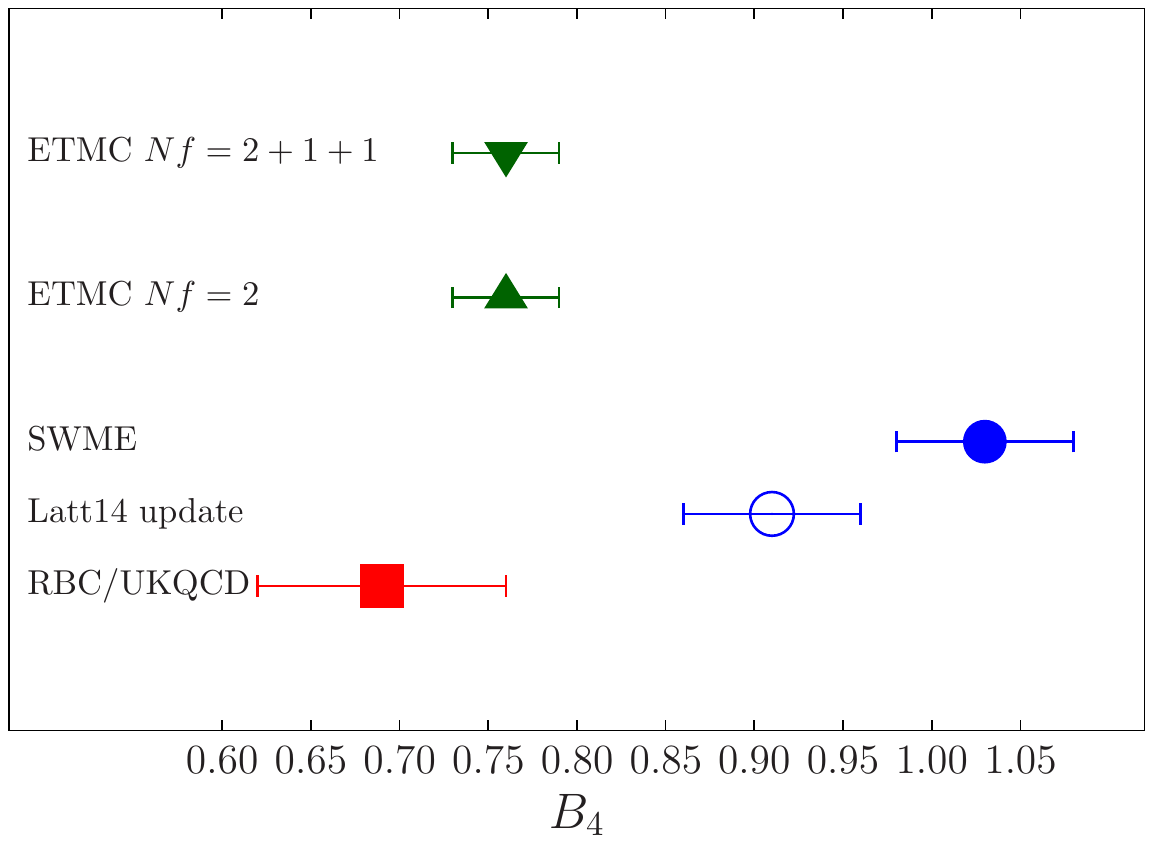} & \hspace*{-3.5cm}\includegraphics[scale=0.34]{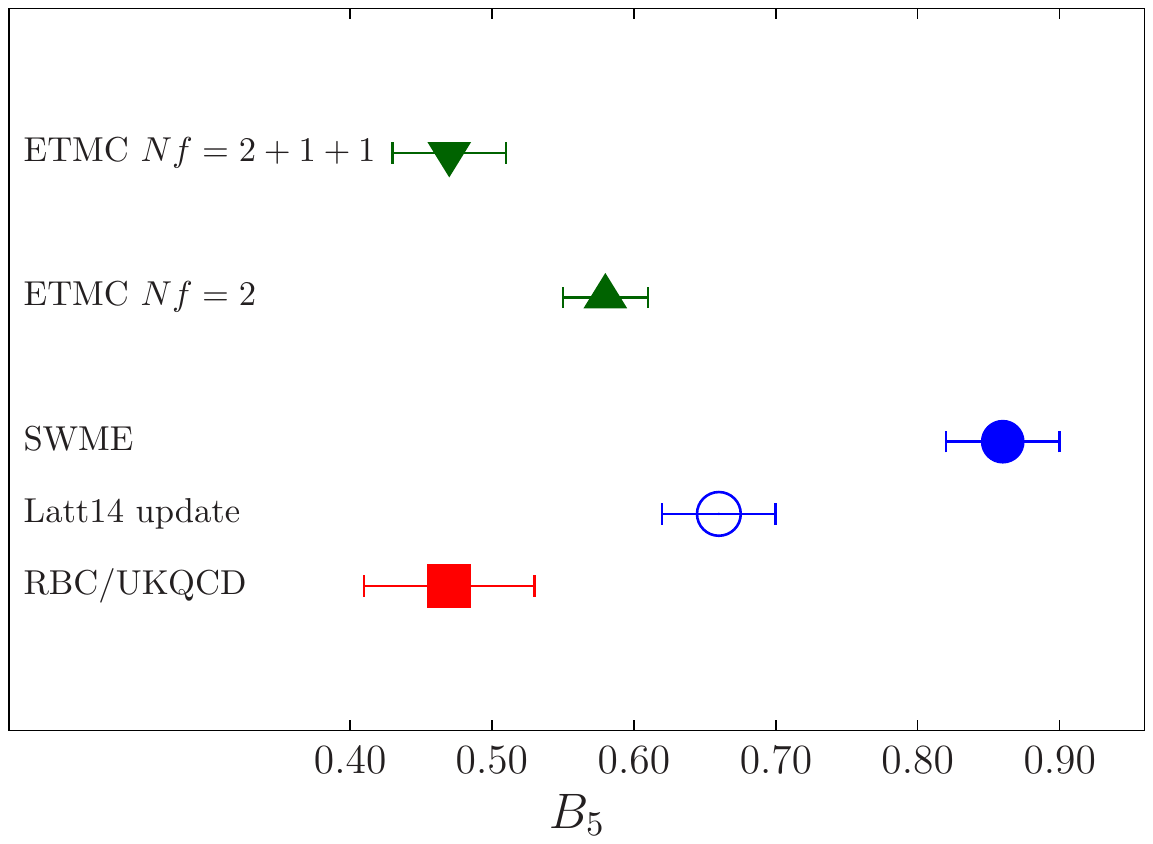}\tabularnewline
\end{tabular}
\vspace*{-5.8cm}

\caption{\label{fig:K-K}$B_{2-5}$ comparison from ETMC (green triangles), SWME (blue circles) and RBC/UKQCD (red squares). For the SWME results, I include the results quoted in  \cite{Bae:2013tca} and the updated results reported at the Lattice 2014 conference  (open circles). }
\end{figure}

Fig.\ref{fig:K-K} and Table \ref{tab:K-K} compare the $B_i$ estimates obtained by the three collaborations. 
As it can be seen in Fig.\ref{fig:K-K}, there is a  $2-3\sigma$   discrepancies for $B_{4}$ and $B_{5}$ between SWME with respect to  ETMC and RBC/UKQC. Further investigation is needed to resolve this discrepancy. As mentioned in \cite{Bae:2013tca}, one possibility is that the truncation errors in the perturbative matching is larger than the estimates.

\vspace*{-0.7cm}
\section{$D^0-\overline{D}^0$  mixing beyond the Standard Model}
\vspace*{-0.1cm}
Very little has been done concerning the calculation of the $\Delta C = 2$ physical matrix elements with unquenched simulations. 

The ETMC has reported on results obtained with their $N_f=2$ ensembles \cite{Carrasco:2014uya} using the same lattice setup as in the $K^0-\overline{K}^0$ analysis \cite{Bertone:2012cu}. Preliminary results with $N_f=2+1+1$ dynamical sea quarks have been also reported in \cite{Carrasco:2013jaa}. The results are compared in Table \ref{tab:D-D}.

There is also an ongoing project by FNAL/MILC reported in \cite{Chang:2013gla} to compute the complete set of D-meson mixing matrix elements on the $N_f=2+1$ Asqtad ensembles with Fermilab clover charm quarks. 

\begin{table}[h!]
\center
\begin{footnotesize}
\begin{tabular}{llccccc}
\hline 
  & $N_f$  \hspace*{-.3cm}  & $B_{1}$ \hspace*{-.25cm} & $B_{2}$ \hspace*{-.25cm} & $B_{3}$ \hspace*{-.25cm}& $B_{4}$\hspace*{-.25cm} & $B_{5}$\hspace*{-.25cm}\tabularnewline
\hline 
ETMC \cite{Carrasco:2014uya}\hspace*{-.3cm} & 2\hspace*{-.3cm} & 0.75(2)\hspace*{-.25cm} & 0.66(2)\hspace*{-.25cm} & 0.97(5) \hspace*{-.25cm}& 0.91(4)\hspace*{-.25cm} & 1.10(5)\hspace*{-.25cm}\tabularnewline
ETMC \cite{tobepublished}\hspace*{-.3cm} & 2+1+1 \hspace*{-.3cm}  & 0.76(4)\hspace*{-.25cm} & 0.64(2)\hspace*{-.25cm} & 1.02(7)\hspace*{-.25cm} & 0.92(3)\hspace*{-.25cm} & 0.95(5)\hspace*{-.25cm}\tabularnewline
\hline 
\end{tabular}
\end{footnotesize}
\caption{\label{tab:D-D}$B_{2-5}$ renormalized in $\overline{MS}$ scheme of \cite{Buras:2000if}  at 3 GeV for D-mixing  obtained by ETMC with $N_f=2$ and $N_f=2+1+1$}
\end{table}

\vspace*{-0.6cm}
\section{Neutral $B_{(s)}$ mixing in the Standard Model}
For heavy quarks, discretization errors grow as $\alpha^k (am_h)^n$ with $k>0$ and $n>0$. With the currently available lattice spacing ($a^{-1} \lesssim 4$ GeV) charm quarks satisfy $am_c \simeq 0.15$ so in general they can use light quark methods if the action is sufficiently improved. For b-quarks, since $am_b \gtrsim 1$, they can not be simulated just like light quarks even with the smallest lattice spacing currently available. In order to circumvent this problem several methods have been proposed based either on implementing an improved light quark action on the lattice and extrapolate to the b-quark mass or in using an effective field theory (EFT)\footnote{for a detailed description of the heavy-quark methods used in lattice QCD see Appendix A.1.3 of \cite{Aoki:2013ldr}.}. To avoid $\mathcal{O}(am_h)$ errors one can discretize a continuum EFT like HQET or NRQCD. Alternatively, the so-called relativistic heavy quark actions (for instance the Fermilab action implemented by FNAL/MILC for heavy quakrs) extend the Symanzick improvement by allowing the coefficients to depend explicitly on $m_h$.  As described in \cite{Christ:2006us} only three parameters ($m_0$, $\eta$ and $c_P$) need to be tunned to remove all discretization errors $\mathcal{O}(am_h)^n$.

Three collaborations have presented results for the B-meson mixing parameters in the SM using $N_f=2+1$ dynamical sea quarks. The HPQCD \cite{Gamiz:2009ku} and FNAL/MILC \cite{Bazavov:2012zs} computations rely on the same $N_f=2+1$ MILC ensembles using staggered AsqTad light valence quarks but while the HPQCD uses nonrelativistic NRQCD action for valence b-quarks \cite{Lepage:1992tx}, FNAL/MILC follows the Fermilab method for b-quarks \cite{ElKhadra:1996mp}.  In  \cite{Gamiz:2009ku}, HPQCD presents results the quantities: $B_{Bs(d)}$, $B_{Bs}/B_{Bd}$, $f_{Bs(d)}\sqrt{B_{Bs(d)}}$  and $\xi$. This computation includes two lattice spacings with $a\sim0.09,0.12$ fm and a minimum pion mass of about 400 MeV. In \cite{Bazavov:2012zs}, the FNAL/MILC collaboration published results for the SU(3) breaking quantities $B_{Bs}/B_{Bd}$ and $\xi$. Preliminary results for the $B_{s}$ and $B_{d}$ mixing quantities $f_{Bs(d)}\sqrt{B_{Bs(d)}}$ and $B_{Bs(d)}$ are presented in \cite{Bouchard:2011xj}.  The calculations in \cite{Bazavov:2012zs,Bouchard:2011xj} include two lattice spacings $a\sim 0.09,0.12$ fm and pion masses as low as 320 MeV. 

Both HPQCD and FNAL/MILC calculate the operator renormalization and matching using one loop improved lattice perturbation theory. This turns out to be the dominant source of systematic uncertainty for $B_{Bs}$ and $B_{Bd}$ while does not result in a significant source of uncertainty for the SU(3) ratios where the statistical, chiral extrapolations and "wrong spin contributions" are the dominant source of uncertainty. 

The so-called "wrong spin contributions", relevant for staggered light quarks, are introduced in \cite{Bazavov:2012zs}. They originate from the mixing of the operator $Q_1$ with the operators $Q_{2-3}$ induced by the interactions between different "tastes". 
This mixing can be accounted for in the chiral and continuum extrapolation by performing a simultaneous fit of the three operators guided by the  staggered chiral perturbation theory proposed in \cite{Bernard:2013dfa} where these contributions have been computed at NLL. The "wrong spin" terms are not included in the chiral extrapolations of \cite{Gamiz:2009ku, Bazavov:2012zs}. However in \cite{Bazavov:2012zs} wrong spin contributions are treated as a systematic error and estimated using the matrix elements of the $Q_{2-3}$ operators in \cite{Bouchard:2011xj}. The estimated uncertainty due to this effect in $\xi$ is 3\% which is the dominant source of uncertainty. Updates from FNAL/MILC with more ensembles, four lattice spacings ranging from $a\sim0.045$ fm to $a\sim0.12$ fm, lighter pion masses and the inclusion of "wrong spin contributions" are reported at the Lattice 2012 \cite{Freeland:2012kz},   2013 \cite{Chang:2013gla} and  2014  conferences. 

Another calculation of the $B$-mixing quantities is presented by the RBC/UKQCD collaboration using the static limit action \cite{Eichten:1989zv} on $N_f=2+1$ domain wall ensembles. 
For many quantities, the corrections $1/m_h$ are estimated to be $\mathcal{O}(\lambda_{QCD}/m_b)\sim 10\%$  and they are needed for a precision calculation. Instead, for ratio quantities like $\xi$,  the static limit can be competitive since the error from the static approximation is reduced to 2\%.  In \cite{Albertus:2010nm} a result for the SU(3) breaking ratio $\xi$ at a single lattice spacing $a\sim0.11$ fm and with a minimum pion mass of 430 MeV is presented. This result has been recently updated in \cite{Aoki:2014nga} with two lattice spacings $a\sim 0.086,0.11$ fm and smaller pion masses ranging from 290 to 420 MeV. The authors of \cite{Aoki:2014nga} also calculate $B_{Bs(d)}$, $f_{Bs(d)}\sqrt{B_{Bs(d)}}$ and $B_{Bs}/B_{Bd}$.  
For the matching between the HQET operators onto the QCD ones, one loop perturbative matching is implemented.
The one-loop renormalization error is estimated to be smaller than 1\% for ratio quantities and around 6\% for non ratio  quantities as $B$-parameters. For this reason, the non-perturbative renormalization is within the future perspectives announced in \cite{Aoki:2014nga}. 

The ETMC has published results for $B_{Bs(d)}$, $\xi$ and $f_{Bs(d)}\sqrt{B_{Bs(d)}}$ in \cite{Carrasco:2013zta} using  $N_f=2$ gauge configuration ensembles. The mixed action fermionic setup in \cite{Bertone:2012cu,Carrasco:2014uya}  is also adopted here and non perturbatively renormalization of four-fermion operators in the RI-MOM scheme is implemented. The extrapolation in the heavy quark mass has been carried out following the ratio method  proposed in \cite{Blossier:2009hg,Dimopoulos:2011gx}, by introducing suitable ratios with exactly known static limit and interpolating them to the b-quark mass. 
In this approach, HQET-QCD perturbative matching factors are only used as an intermediate step to construct the ratios with the correct HQET scaling law.

\begin{figure}[h!]
 \centering
 \vspace*{-2.3cm}
 \hspace*{-2.0cm}\includegraphics[scale=0.5]{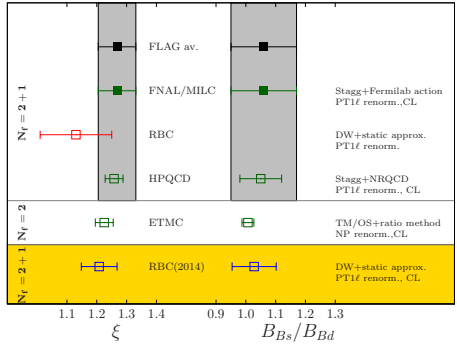}
 \vspace*{-7.9cm}

 \caption{ \label{fig:xi}Overview of lattice results for the SU(3) ratios $\xi$ and $B_{Bs}/B_{Bd}$. FLAG averages have been also added. The colour code is the same as in Fig.\ref{fig:BK}.}
\end{figure}

\begin{figure}[h!]
 \centering
 \vspace*{-2.8cm}
 \hspace*{-2.0cm}\includegraphics[scale=0.5]{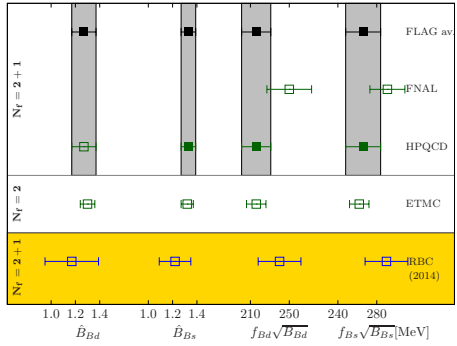}
 \vspace*{-7.9cm}
 \caption{ \label{fig:B}Overview of lattice results for the quantities $\hat{B}_{Bs}$, $\hat{B}_{Bd}$, $f_{Bd}\sqrt{B_{Bd}}$ and $f_{Bs}\sqrt{B_{Bs}}$. The colour code is the same as in Fig.\ref{fig:BK}.}
\end{figure}

A comparison of all $N_f=2$ and $N_f=2+1$ results is shown in Figs.\ref{fig:xi} and \ref{fig:B}. Fig.\ref{fig:xi} shows a comparison of the results for the SU(3) breaking ratios $\xi$ and $B_{Bs}/B_{Bd}$ while Fig.\ref{fig:B} collects the RGI SM $B$-parameters and the quantity  $f_B\sqrt{B_B}$ in MeV for $B_d$ and $B_s$ mesons. These results are summarized in Tables \ref{tab:xi} and \ref{tab:B}.   In both figures and in both tables  the systematic error due to the static approximation in the RBC computation, estimated in 2\% for ratio quantities and 10\% for the rest, is not included in the error.

\vspace*{-0.4cm}
\begin{table}[h!]
\center
\begin{footnotesize}
\begin{tabular}{llll}
\hline 
Collaboration  \hspace*{-0.35cm} &   $N_{f}$ &  \hspace*{0.2cm}$\xi$  &   \hspace*{-0.2cm}$B_{Bs}/B_{Bd}$  \tabularnewline
\hline 
ETMC \cite{Carrasco:2013zta} \hspace*{-.4cm}  & 2 & \hspace*{-0.3cm} 1.225(30) & \hspace*{-0.3cm} 1.007(20) \tabularnewline
FNAL \cite{Bazavov:2012zs}  \hspace*{-.4cm}  & 2+1 & \hspace*{-0.3cm} 1.268(63)  & \hspace*{-0.3cm}  1.06(11)\tabularnewline
RBC/UKQCD \cite{Albertus:2010nm} \hspace*{-.4cm}  & 2+1 & \hspace*{-0.3cm}   1.13(12) & \hspace*{0.3cm} - \tabularnewline
HPQCD \cite{Gamiz:2009ku} \hspace*{-.4cm}  & 2+1 & \hspace*{-0.3cm}   1.258(33)&  \hspace*{-0.3cm}  1.05(7)\tabularnewline
RBC  \cite{Aoki:2014nga} \hspace*{-.4cm}  & 2+1 & \hspace*{-0.3cm} 1.208(60) & \hspace*{-0.3cm}  1.028(74) \tabularnewline
FLAG  \cite{Aoki:2013ldr}\hspace*{-.4cm}  & 2+1 & \hspace*{-0.3cm} 1.268(63) & \hspace*{-0.3cm}  1.06(11) \tabularnewline
\hline

\end{tabular}
\end{footnotesize}
\caption{\label{tab:xi}Recent lattice results for $\xi$ and $B_{Bs}/B_{Bd}$ by the different collaborations and the FLAG average.}
\end{table}

\vspace*{-0.5cm}
\begin{table}[h!]
\center
\begin{footnotesize}

\begin{tabular}{llllll}
\hline 
Collaboration \hspace*{-.4cm} &   $N_{f}$ \hspace*{-0.3cm} &    \hspace*{-0.4cm}$\hat{B}_{Bd}$   &   \hspace*{-0.4cm} $\hat{B}_{Bs}$   &   \hspace*{-0.4cm} $f_{Bd}\sqrt{B_{Bd}}$   &   \hspace*{-0.4cm} $f_{Bs}\sqrt{B_{Bs}}$ \tabularnewline
\hline 
ETMC \cite{Carrasco:2013zta} \hspace*{-.4cm}  & 2 & \hspace*{-0.4cm} 1.30(6)  & \hspace*{-0.4cm} 1.32(5) & \hspace*{-0.4cm} 216(10) & \hspace*{-0.4cm} 262(10)\tabularnewline
FNAL \cite{Bazavov:2012zs}  \hspace*{-.4cm}  & 2+1 & \hspace*{-0.4cm}   & \hspace*{-0.4cm} & \hspace*{-0.4cm} 250(23)  & \hspace*{-0.4cm} 291(18)\tabularnewline
HPQCD \cite{Gamiz:2009ku} \hspace*{-.4cm}  & 2+1 & \hspace*{-0.4cm} 1.27(10)  &  \hspace*{-0.4cm} 1.33(6) &  \hspace*{-0.4cm} 216(15)  & \hspace*{-0.4cm} 266(18) \tabularnewline
RBC  \cite{Aoki:2014nga} \hspace*{-.4cm}  & 2+1 & \hspace*{-0.4cm} 1.17(22) & \hspace*{-0.4cm} 1.22(13)  & \hspace*{-0.4cm} 240(22)   & \hspace*{-0.4cm} 290(22) \tabularnewline
FLAG \cite{Aoki:2013ldr} \hspace*{-.4cm}  & 2+1 & \hspace*{-0.4cm} 1.27(10)  & \hspace*{-0.4cm} 1.33(6) & \hspace*{-0.4cm} 216(15) & \hspace*{-0.4cm} 266(18)\tabularnewline
\hline

\end{tabular}
\end{footnotesize}
\caption{\label{tab:B}Recent lattice results for   RGI $B_{d(s)}$ and $f_{Bd(s)}\sqrt{B_{Bd(s)}}$ in MeV by the different collaborations and the FLAG average. }
\end{table}
\vspace*{-1.0cm}
\section{$B^0-\overline{B}^0$  mixing beyond the Standard Model}
The ETMC has computed the non SM $B_{2-5}$ parameters employing the ratio method approach on $N_f=2$ ensembles  \cite{Carrasco:2013zta}. There is also  work in progress by FNAL/MILC to compute the $B$-parameters complete basis of four-fermion operators. Preliminary results can be found in  \cite{Bouchard:2011xj} and updates have been reported at the Lattice 2012 \cite{Freeland:2012kz}, 2013 \cite{Chang:2013gla} and 2014 conferences. Table \ref{tab:B-B} reports both results.  Note that  FNAL/MILC $B_{2-5}$ parameters defined in Eq.(1.4) of \cite{Bouchard:2011xj}  differ from the ETMC definition, which is the one  in Eq.(\ref{eq:Bi-def}), by a factor of\footnote{I would like to thank Elvira Gamiz for bringing to my attention this point.}  $M^2_{B(s)}/(m_b(\mu)+m_q(\mu))^ 2$ . 

\begin{table}[h!]
\center
\begin{footnotesize}
\begin{tabular}{lccccc}
\hline 
 ETMC $i=$\hspace*{-0.35cm} & 1\hspace*{-0.35cm} & 2\hspace*{-0.35cm}& 3 \hspace*{-0.35cm}& 4\hspace*{-0.35cm} & 5\hspace*{-0.35cm} \tabularnewline
\hline
$f_{Bd}\sqrt{B^{(d)}_i}$ \hspace*{-0.35cm} & 174(8) \hspace*{-0.35cm}& 160(8)\hspace*{-0.35cm} & 177(17)\hspace*{-0.35cm} & 185(9)\hspace*{-0.35cm} & 229(14)\hspace*{-0.35cm} \tabularnewline
$f_{Bs}\sqrt{B^{(s)}_i}$ \hspace*{-0.35cm} & 211(8) \hspace*{-0.35cm}& 195(7)\hspace*{-0.35cm} & 215(17)\hspace*{-0.35cm} & 220(9)\hspace*{-0.35cm} & 285(14)\hspace*{-0.35cm} \tabularnewline
\hline

\hline
FNAL/MILC $i=$\hspace*{-0.35cm} & 1\hspace*{-0.35cm} & 2\hspace*{-0.35cm}& 3 \hspace*{-0.35cm}& 4\hspace*{-0.35cm} & 5\hspace*{-0.35cm} \tabularnewline
\hline
$f_{Bd}\sqrt{B^{(d)}_i}$ \hspace*{-0.35cm}&  202(36) \hspace*{-0.35cm}& 183(11) \hspace*{-0.35cm}& 190(36)\hspace*{-0.35cm} & 241(26)\hspace*{-0.35cm} & 282(33)\hspace*{-0.35cm} \tabularnewline
$f_{Bs}\sqrt{B^{(s)}_i}$ \hspace*{-0.35cm}&  236(29) \hspace*{-0.35cm}& 225(28) \hspace*{-0.35cm}& 231(38)\hspace*{-0.35cm} & 293(32)\hspace*{-0.35cm} & 336(38)\hspace*{-0.35cm} \tabularnewline
\hline
\end{tabular}
\end{footnotesize}
\caption{ \label{tab:B-B}$f_B\sqrt{B_i}$ results from ETMC \cite{Carrasco:2013zta} and  FNAL/MILC \cite{Bouchard:2011xj} for $B_d$ and $B_s$ mesons. $B_i$ are renormalized in $\overline{MS}$ scheme of \cite{Buras:2000if} at the scale of b-quark mass. Due to the different definition of $B_{2-5}$ between ETMC and FNAL/MILC, for the comparison, the FNAL/MILC results have been converted to the definition in Eq.(\ref{eq:Bi-def}) using the factor $M_{B(s)}/(m_b(\mu)+m_q(\mu))$ with the PDG values for $m_b$ and $m_q$.}
\end{table}

\vspace*{-0.8cm}

\section{Deviation from  VIA}
Finally, it is interesting to mention the results reported in \cite{Carrasco:2013jda} where the deviation of the VIA for the matrix elements of the complete basis of $\Delta F = 2$ operators are systematically investigated.  Large violations of the VIA are found in the kaon sector, in particular for one of the two relevant Wick contractions which confirm the results found in  \cite{Boyle:2012ys} for the operator contributing in the SM. These deviations decrease  as the meson mass increases and the VIA predictions turn out to provide closer results to the lattice ones for B-mesons and, even better, in the infinite mass limit. 
\vspace*{-0.4cm}
\section{Conclusions}
In the last years, several  collaborations using different lattice regularizations  have provided new or updated unquenched results for the $B_K$ parameter that are in nice agreement between them. The comparison of $B_K$ results indicate that the systematic error introduced by the quenching is smaller than other systematic uncertainties.  
The $B_K$ estimate from the lattice, which historically was the larger source of uncertainty in the $\epsilon_K$ determination of Eq.(\ref{eq:epsilon_K}), is now in the third place in the error budget.  The main sources of systematic error in the $B_K$ computation used to be the chiral extrapolation to the physical point and the renormalization procedure. Nowadays, simulations with near physical pion masses are feasible and many collaborations make use of non perturbative renormalization. As a consequence, the $B_K$ computation has entered in the era of precision measurements with total estimated uncertainty of 1.5-4\%, with the largest systematic error coming in most cases from the  renormalization. 

In contrast, physical pion mass gauge ensembles are still not used for mixing quantities in the $B$-sector. Of course, $B$-physics is a very active field in the lattice community and  a significant improvement for these quantities is expected in the next few years. 
 
Since there are several groups using different lattice methods to calculate the same quantities and provide complete error budgets, there is the necessity of combining all these results in a final number. 
This is actually the scope of the FLAG collaboration, that is, to provide the best lattice estimate to be used in a phenomenological analysis. In particular, this is the case for the $K$- and $B$-mixing parameters in the SM.
 
There is also a recent effort to compute the matrix elements of the full $\Delta F = 2 $ operators. Further investigations are needed here since in the $K^{0}-\overline{K}^{0}$ mixing RBC/UKQCD and ETMC have found compatible results, but $B_{4-5}$ values are in tension with the ones obtained by SWME. For $f_{B}\sqrt{B_{i}}$, the FNAL/MILC and ETMC results have discrepancies at the level of 1-2$\sigma$ for $B_2$,$B_3$,$B_5$ and 3$\sigma$ for $B_{4}$. 

$B$-parameters BSM are still not included among the FLAG averages.   ETMC is the only collaboration up to now that has published final results for the $B_{2-5}$ parameters in the $K$, $D$ and $B$ sector providing a complete error budget. These results can be used as input in the relevant phenomenological analyses.

\vspace*{-0.45cm}
\section{Acknowledgements}

I wish to thank the organizers of the ICHEP2014 conference for the invitation to present this review. I am grateful to Petros Dimopoulos, Vittorio Lubicz, Silvano Simula and Cecilia Tarantino for useful discussions and a careful reading of the proceeding. 
\vspace*{-0.5cm}



\nocite{*}
\bibliographystyle{elsarticle-num}
\bibliography{biblio}

\begin{thebibliography}{5}
\scriptsize
\expandafter\ifx\csname url\endcsname\relax
  \def\url#1{\texttt{#1}}\fi
\expandafter\ifx\csname urlprefix\endcsname\relax\def\urlprefix{URL }\fi
\expandafter\ifx\csname href\endcsname\relax
  \def\href#1#2{#2} \def\path#1{#1}\fi

\bibitem{Buras:1990fn}
A.~J. Buras, et~al., Nucl.Phys. B347 (1990) 491--536.
\newblock \href {http://dx.doi.org/10.1016/0550-3213(90)90373-L}
  {\path{doi:10.1016/0550-3213(90)90373-L}}.

\bibitem{Herrlich:1993yv}
S.~Herrlich, U.~Nierste, Nucl.Phys. B419 (1994) 292--322.
\newblock \href {http://arxiv.org/abs/hep-ph/9310311}
  {\path{arXiv:hep-ph/9310311}}, \href
  {http://dx.doi.org/10.1016/0550-3213(94)90044-2}
  {\path{doi:10.1016/0550-3213(94)90044-2}}.

\bibitem{Herrlich:1996vf}
S.~Herrlich, U.~Nierste, Nucl.Phys. B476 (1996) 27--88.
\newblock \href {http://arxiv.org/abs/hep-ph/9604330}
  {\path{arXiv:hep-ph/9604330}}, \href
  {http://dx.doi.org/10.1016/0550-3213(96)00324-0}
  {\path{doi:10.1016/0550-3213(96)00324-0}}.

\bibitem{Brod:2010mj}
J.~Brod, M.~Gorbahn, Phys.Rev. D82 (2010) 094026.
\newblock \href {http://arxiv.org/abs/1007.0684} {\path{arXiv:1007.0684}},
  \href {http://dx.doi.org/10.1103/PhysRevD.82.094026}
  {\path{doi:10.1103/PhysRevD.82.094026}}.

\bibitem{Brod:2011ty}
J.~Brod, M.~Gorbahn, Phys.Rev.Lett. 108 (2012) 121801.
\newblock \href {http://arxiv.org/abs/1108.2036} {\path{arXiv:1108.2036}},
  \href {http://dx.doi.org/10.1103/PhysRevLett.108.121801}
  {\path{doi:10.1103/PhysRevLett.108.121801}}.

\bibitem{Ciuchini:1997bw}
M.~Ciuchini, et~al., Nucl.Phys. B523 (1998) 501--525.
\newblock \href {http://arxiv.org/abs/hep-ph/9711402}
  {\path{arXiv:hep-ph/9711402}}, \href
  {http://dx.doi.org/10.1016/S0550-3213(98)00161-8}
  {\path{doi:10.1016/S0550-3213(98)00161-8}}.

\bibitem{Bertone:2012cu}
V.~Bertone, et~al., JHEP 1303 (2013) 089.
\newblock \href {http://arxiv.org/abs/1207.1287} {\path{arXiv:1207.1287}},
  \href {http://dx.doi.org/10.1007/JHEP07(2013)143, 10.1007/JHEP03(2013)089}
  {\path{doi:10.1007/JHEP07(2013)143, 10.1007/JHEP03(2013)089}}.

\bibitem{Carrasco:2014uya}
N.~Carrasco, et~al., Phys.Rev. D90 (2014) 014502.
\newblock \href {http://arxiv.org/abs/1403.7302} {\path{arXiv:1403.7302}},
  \href {http://dx.doi.org/10.1103/PhysRevD.90.014502}
  {\path{doi:10.1103/PhysRevD.90.014502}}.

\bibitem{Carrasco:2013zta}
N.~Carrasco, et~al., JHEP 1403 (2014) 016.
\newblock \href {http://arxiv.org/abs/1308.1851} {\path{arXiv:1308.1851}},
  \href {http://dx.doi.org/10.1007/JHEP03(2014)016}
  {\path{doi:10.1007/JHEP03(2014)016}}.

\bibitem{Frezzotti:2004wz}
R.~Frezzotti, G.~Rossi, JHEP 0410 (2004) 070.
\newblock \href {http://arxiv.org/abs/hep-lat/0407002}
  {\path{arXiv:hep-lat/0407002}}, \href
  {http://dx.doi.org/10.1088/1126-6708/2004/10/070}
  {\path{doi:10.1088/1126-6708/2004/10/070}}.

\bibitem{Aoki:2007xm}
Y.~Aoki, et~al., Phys.Rev. D78 (2008) 054510.
\newblock \href {http://arxiv.org/abs/0712.1061} {\path{arXiv:0712.1061}},
  \href {http://dx.doi.org/10.1103/PhysRevD.78.054510}
  {\path{doi:10.1103/PhysRevD.78.054510}}.

\bibitem{Durr:2011ap}
S.~Durr, et~al., Phys.Lett. B705 (2011) 477--481.
\newblock \href {http://arxiv.org/abs/1106.3230} {\path{arXiv:1106.3230}},
  \href {http://dx.doi.org/10.1016/j.physletb.2011.10.043}
  {\path{doi:10.1016/j.physletb.2011.10.043}}.

\bibitem{Bae:2014sja}
T.~Bae, et~al., Phys.Rev. D89 (2014) 074504.
\newblock \href {http://arxiv.org/abs/1402.0048} {\path{arXiv:1402.0048}},
  \href {http://dx.doi.org/10.1103/PhysRevD.89.074504}
  {\path{doi:10.1103/PhysRevD.89.074504}}.

\bibitem{Laiho:2011np}
J.~Laiho, R.~S. Van~de Water, PoS LATTICE2011 (2011) 293.
\newblock \href {http://arxiv.org/abs/1112.4861} {\path{arXiv:1112.4861}}.

\bibitem{Arthur:2012opa}
R.~Arthur, et~al., Phys.Rev. D87 (2013) 094514.
\newblock \href {http://arxiv.org/abs/1208.4412} {\path{arXiv:1208.4412}},
  \href {http://dx.doi.org/10.1103/PhysRevD.87.094514}
  {\path{doi:10.1103/PhysRevD.87.094514}}.

\bibitem{tobepublished}
ETMC, paper in preparation.

\bibitem{Aoki:2013ldr}
FLAG, S.~Aoki, et~al.\href {http://arxiv.org/abs/1310.8555}
  {\path{arXiv:1310.8555}}.

\bibitem{Bae:2011ff}
T.~Bae, et~al., Phys.Rev.Lett. 109 (2012) 041601.
\newblock \href {http://arxiv.org/abs/1111.5698} {\path{arXiv:1111.5698}},
  \href {http://dx.doi.org/10.1103/PhysRevLett.109.041601}
  {\path{doi:10.1103/PhysRevLett.109.041601}}.

\bibitem{Aubin:2009jh}
C.~Aubin, J.~Laiho, R.~S. Van~de Water, Phys.Rev. D81 (2010) 014507.
\newblock \href {http://arxiv.org/abs/0905.3947} {\path{arXiv:0905.3947}},
  \href {http://dx.doi.org/10.1103/PhysRevD.81.014507}
  {\path{doi:10.1103/PhysRevD.81.014507}}.

\bibitem{Frison:2013fga}
J.~Frison, et~al., PoS LATTICE2013 (2013) 460.
\newblock \href {http://arxiv.org/abs/1312.2374} {\path{arXiv:1312.2374}}.

\bibitem{Constantinou:2010qv}
M.~Constantinou, et~al., Phys.Rev. D83 (2011) 014505.
\newblock \href {http://arxiv.org/abs/1009.5606} {\path{arXiv:1009.5606}},
  \href {http://dx.doi.org/10.1103/PhysRevD.83.014505}
  {\path{doi:10.1103/PhysRevD.83.014505}}.

\bibitem{Carrasco:2013jaa}
N.~Carrasco, et~al., PoS LATTICE2013 (2013) 393.
\newblock \href {http://arxiv.org/abs/1310.5461} {\path{arXiv:1310.5461}}.

\bibitem{Bae:2013tca}
T.~Bae, et~al., Phys.Rev. D88~(7) (2013) 071503.
\newblock \href {http://arxiv.org/abs/1309.2040} {\path{arXiv:1309.2040}},
  \href {http://dx.doi.org/10.1103/PhysRevD.88.071503}
  {\path{doi:10.1103/PhysRevD.88.071503}}.

\bibitem{Boyle:2012qb}
P.~Boyle, N.~Garron, R.~Hudspith, Phys.Rev. D86 (2012) 054028.
\newblock \href {http://arxiv.org/abs/1206.5737} {\path{arXiv:1206.5737}},
  \href {http://dx.doi.org/10.1103/PhysRevD.86.054028}
  {\path{doi:10.1103/PhysRevD.86.054028}}.

\bibitem{Lytle:2013oqa}
A.~Lytle, et~al., PoS LATTICE2013 (2013) 400.
\newblock \href {http://arxiv.org/abs/1311.0322} {\path{arXiv:1311.0322}}.

\bibitem{Buras:2000if}
A.~J. Buras, et~al., Nucl.Phys. B586 (2000) 397--426.
\newblock \href {http://arxiv.org/abs/hep-ph/0005183}
  {\path{arXiv:hep-ph/0005183}}, \href
  {http://dx.doi.org/10.1016/S0550-3213(00)00437-5}
  {\path{doi:10.1016/S0550-3213(00)00437-5}}.

\bibitem{Chang:2013gla}
C.~Chang, et~al., PoS LATTICE2013  477.
\newblock \href {http://arxiv.org/abs/1311.6820} {\path{arXiv:1311.6820}}.

\bibitem{Christ:2006us}
N.~H. Christ, M.~Li, H.-W. Lin, Phys.Rev. D76 (2007) 074505.
\newblock \href {http://arxiv.org/abs/hep-lat/0608006}
  {\path{arXiv:hep-lat/0608006}}, \href
  {http://dx.doi.org/10.1103/PhysRevD.76.074505}
  {\path{doi:10.1103/PhysRevD.76.074505}}.

\bibitem{Gamiz:2009ku}
E.~Gamiz, et~al., Phys.Rev. D80 (2009) 014503.
\newblock \href {http://arxiv.org/abs/0902.1815} {\path{arXiv:0902.1815}},
  \href {http://dx.doi.org/10.1103/PhysRevD.80.014503}
  {\path{doi:10.1103/PhysRevD.80.014503}}.

\bibitem{Bazavov:2012zs}
A.~Bazavov, et~al., Phys.Rev. D86 (2012) 034503.
\newblock \href {http://arxiv.org/abs/1205.7013} {\path{arXiv:1205.7013}},
  \href {http://dx.doi.org/10.1103/PhysRevD.86.034503}
  {\path{doi:10.1103/PhysRevD.86.034503}}.

\bibitem{Lepage:1992tx}
G.~e.~a. Lepage, Phys.Rev. D46 (1992) 4052--4067.
\newblock \href {http://arxiv.org/abs/hep-lat/9205007}
  {\path{arXiv:hep-lat/9205007}}, \href
  {http://dx.doi.org/10.1103/PhysRevD.46.4052}
  {\path{doi:10.1103/PhysRevD.46.4052}}.

\bibitem{ElKhadra:1996mp}
A.~X. e.~a. El-Khadra, Phys.Rev. D55 (1997) 3933--3957.
\newblock \href {http://arxiv.org/abs/hep-lat/9604004}
  {\path{arXiv:hep-lat/9604004}}, \href
  {http://dx.doi.org/10.1103/PhysRevD.55.3933}
  {\path{doi:10.1103/PhysRevD.55.3933}}.

\bibitem{Bouchard:2011xj}
C.~Bouchard, et~al., PoS LATTICE2011 (2011) 274.
\newblock \href {http://arxiv.org/abs/1112.5642} {\path{arXiv:1112.5642}}.

\bibitem{Bernard:2013dfa}
C.~Bernard, Phys.Rev. D87~(11) (2013) 114503.
\newblock \href {http://arxiv.org/abs/1303.0435} {\path{arXiv:1303.0435}},
  \href {http://dx.doi.org/10.1103/PhysRevD.87.114503}
  {\path{doi:10.1103/PhysRevD.87.114503}}.

\bibitem{Freeland:2012kz}
E.~Freeland, et~al., PoS LATTICE2012 (2012) 124.
\newblock \href {http://arxiv.org/abs/1212.5470} {\path{arXiv:1212.5470}}.

\bibitem{Eichten:1989zv}
E.~Eichten, B.~R. Hill, Phys.Lett. B234 (1990) 511.
\newblock \href {http://dx.doi.org/10.1016/0370-2693(90)92049-O}
  {\path{doi:10.1016/0370-2693(90)92049-O}}.

\bibitem{Albertus:2010nm}
C.~Albertus, et~al., Phys.Rev. D82 (2010) 014505.
\newblock \href {http://arxiv.org/abs/1001.2023} {\path{arXiv:1001.2023}},
  \href {http://dx.doi.org/10.1103/PhysRevD.82.014505}
  {\path{doi:10.1103/PhysRevD.82.014505}}.

\bibitem{Aoki:2014nga}
Y.~Aoki, et~al.\href {http://arxiv.org/abs/1406.6192} {\path{arXiv:1406.6192}}.

\bibitem{Blossier:2009hg}
B.~Blossier, et~al., JHEP 1004 (2010) 049.
\newblock \href {http://arxiv.org/abs/0909.3187} {\path{arXiv:0909.3187}},
  \href {http://dx.doi.org/10.1007/JHEP04(2010)049}
  {\path{doi:10.1007/JHEP04(2010)049}}.

\bibitem{Dimopoulos:2011gx}
P.~Dimopoulos, et~al., JHEP 1201 (2012) 046.
\newblock \href {http://arxiv.org/abs/1107.1441} {\path{arXiv:1107.1441}},
  \href {http://dx.doi.org/10.1007/JHEP01(2012)046}
  {\path{doi:10.1007/JHEP01(2012)046}}.

\bibitem{Carrasco:2013jda}
N.~Carrasco, V.~Lubicz, L.~Silvestrini, $\,$\href
  {http://arxiv.org/abs/1312.6691} {\path{arXiv:1312.6691}}, \href
  {http://dx.doi.org/10.1016/j.physletb.2014.07.016}
  {\path{doi:10.1016/j.physletb.2014.07.016}}.

\bibitem{Boyle:2012ys}
P.~Boyle, et~al., Phys.Rev.Lett. 110~(15) (2013) 152001.
\newblock \href {http://arxiv.org/abs/1212.1474} {\path{arXiv:1212.1474}},
  \href {http://dx.doi.org/10.1103/PhysRevLett.110.152001}
  {\path{doi:10.1103/PhysRevLett.110.152001}}.

\bibitem{Becirevic:2001xt}
D.~Becirevic, et~al., JHEP 0204 (2002) 025.
\newblock \href {http://arxiv.org/abs/hep-lat/0110091}
  {\path{arXiv:hep-lat/0110091}}, \href
  {http://dx.doi.org/10.1088/1126-6708/2002/04/025}
  {\path{doi:10.1088/1126-6708/2002/04/025}}.

\end{thebibliography}







\end{document}